**Evidence for reentrant quantum paraelectric state preceded by a multiglass phase with non-classical exponent and magnetodielectric coupling in SrFe$_{12}$O$_{19}$**

Keshav Kumar[1,2] and Dhananjai Pandey[1,3,4]

[1]School of Materials Science and Technology, Indian Institute of Technology (Banaras Hindu University), Varanasi-221005, India

[2]New Chemistry Unit, International Centre for Materials Science, Jawaharlal Nehru Centre for Advanced Scientific Research, Jakkur, Bengaluru 560064, India

[3]Department of Metallurgical and Materials Engineering, Indian Institute of Technology Bhilai, Raipur-492015, India

[4]Department of Physics, Banaras Hindu University, Varanasi-221005, India

**Abstract**

Evidence for a re-entrant quantum paraelectric (QPE) state preceded by a dipole glass (DG) phase with a non-classical exponent in the quantum critical regime of SrFe$_{12}$O$_{19}$ is presented. It is shown that the DG transition is accompanied with a spin glass (SG) transition and presence of a biquadratic coupling of two diverse order parameter fields. Further, the ergodic symmetry breaking temperatures for the DG and SG transitions coincide (T$_{DG}$ ∼ T$_{SG}$) within ±1K suggesting that SrFe$_{12}$O$_{19}$ exhibits a canonical multiglass state. The stability of the dipole glass state is enhanced magnetically as evidenced by the increase in the freezing temperature with magnetic field (H). The re-entrant QPE state, on the other hand, is found to give way to another frequency dependent peak in the temperature dependence of dielectric constant, most likely a DG phase, at a constant H. Further, this transition is not linked to any magnetic transition in sharp contrast to the higher temperature multiglass transition. The transition temperature of this phase decreases with increasing magnetic field for a fixed frequency unlike the higher temperature DG transition. This raises the possibility of locating a quantum critical point (QCP) in this system at higher magnetic fields than that used in the present work. These results are discussed in the light of quantum critical models of multiferroic transitions. Our results highlight the need for more theoretical studies specific to multiferroic quantum criticality in a multiglass system.

-------------------------------------------------------------------------------------------------

Quantum phase transitions (QPTs) continue to evince considerable interest [1–3] going well beyond its traditional domains of applications to metallic ferromagnetism, heavy fermions, geometrically frustrated 2D antiferromagnets, topological systems and superconductors to other systems like ferroelectrics [4], confined water molecules [5] and quantum computing [6,7]. QPTs occur at the experimentally inaccessible zero temperature (0K) as a result of variation of a non-thermal control parameters (g) like pressure, composition, electric and magnetic fields, which destabilizes the quantum disordered or symmetric state, by suppressing the quantum fluctuations, and stabilises a quantum ordered state with broken symmetry [1]. However, the real interest in the QPT systems lies in the experimentally accessible quantum critical regime (QCR), above the QCP, where both the quantum ($\hbar\omega_0$) and thermal ($k_BT$) fluctuations compete and give rise to various exotic quantum states of matter with potential for technological applications and new physics [1,2]. Unlike the classical systems, the potential and kinetic energy parts of the Hamiltonian do not commute in the quantum systems but may be mapped to a classical system in (d+z) dimension, where z is the dynamical exponent for the dispersion relationship [1]. QPTs have already been shown to have profound impact on several high temperature phenomena such as superconductivity in cuprates [8] and Mott insulators [9].

The studies on the marginal/incipient ferroelectrics (FEs) of displacive type, such as $SrTiO_3$ and $KTaO_3$, which are prevented from entering into a long range ordered FE state even at low temperatures (down to mK) due to quantum fluctuations [10–13] have witnessed a spurt in recent times due to their potential for technological applications [e.g., see 14–17] and the emergence of new physics [e.g., see 18–22]. The dielectric constant $\varepsilon_r$ of both $SrTiO_3$ and $KTaO_3$ [10–13] increases with decreasing temperature as per Curie-Weiss (C-W) law while the square of the TO mode frequency ($\omega_{TO}^2$) decreases linearly with temperature [23–26], consistent with Lyddane-Sachs-Teller (LST) relationship [26], up to a characteristic temperature $T_1$. Below $T_1$, the $\varepsilon_r(T)$ continues to increase but deviates significantly from C-W law. On further lowering of the temperature, $\varepsilon_r(T)$ shows a nearly temperature independent plateau with a very high value of $\varepsilon_r$ in the range of ~24000 to 40000 in SrTiO$_3$ [10,27] and ~4000 in KTaO$_3$ [28]. The $\omega_{TO}^2$ (T) also saturates to a rather low value which never goes to zero in the limit of T $\rightarrow$ 0K [23–25], *suggesting the stabilisation of the paraelectric state.* Muller and Burkhard [10] attributed it to the zero-point fluctuations in the atomic positions of Ti$^{4+}$ in the context of SrTiO$_3$ at low temperatures. They accordingly termed such compounds as quantum paraelectrics (QPE) [10], which may be regarded as being on the disordered side of the FE QCP. They used the quantum mechanical analogue of Curie-Weiss law, commonly known as Barret's law [29], to explain the observed temperature dependence of the dielectric constant and its saturation below 4K in SrTiO$_3$, $\varepsilon = \varepsilon_0 + M/(0.5T_1\coth(T_1/2T) + T_0)$, where $M$ is related to dipole density, $T_0$ is like Curie-Weiss (C-W) temperature and $T_1$ is the crossover temperature from classical to quantum regime. Since then a large number of quantum paraelectrics have been discovered with quantum saturation regime extending to much higher temperatures (up to ~50 K) in some high temperature quantum paraelectrics [30,31]) with both positive [10,32] and negative $T_0$ [30,32–36] corresponding to their incipient FE [13] and incipient antiferroelectric (AFE) [36] characters, respectively. Use of non-thermal control parameters (g), like composition [36–38], electric field [39,40], pressure [12,28], have been shown to drive these compounds to a quantum ferroelectric (QFE) state on the other side of QCP with $T_c$ varying as $(g-g_c)^{1/2}$ with a non-classical exponent (1/2) in the QCR [11,13,37]. The so-called QFE state in Ca$^{2+}$ doped SrTiO$_3$ (SCT) [37] is now known to be a quantum ferrielectric (QFI) state [41] which transforms to an unusual AFE phase at higher Ca$^{2+}$ concentrations [42].

Barrett's law is a single ion model corresponding to the soft TO branch in which each atom behaves as an independent Einstein oscillator in the presence of a small anharmonic term added to the potential energy. Recently, it has been shown [43] that the inverse of the dielectric susceptibility ($\chi_e(T)^{-1}$) of QPEs, like SrTiO$_3$ and KTaO$_3$, in the QCR exhibits an upturn with a T$^2$ dependence above the upturn temperature in disagreement with the dielectric plateau predicted by Barrett's law. Such an upturn is, however, in agreement with the theoretical predictions for the multiaxial displacive FEs in the quantum limit taking into account the coupling of the soft dispersive TO branch with all other phonon branches in the anharmonic term [13,43–48]. Since then other magnetic QPEs, like BaFe$_{12}$O$_{19}$ (BFO) [49] and SrFe$_{12}$O$_{19}$ (SFO) [49], have also been shown to exhibit similar upturns but with $\chi_e(T)^{-1}$ varying as T$^3$ above the upturn temperature, as expected for uniaxial displacive FEs in the quantum limit [44]. More recently, the quantum critical behaviour of multiferroic systems, where the magnetic and FE QCPs can be tuned by some non-thermal control parameters, has also been discussed theoretically [50,51] with interesting predictions such as: (1) the existence of a magnetically enhanced polar state that transforms to a re-entrant QPE state on lowering the temperature at a constant value of the non-thermal control parameter, and (2) the critical exponents for $\chi_e(T)^{-1}$ of such multiferroic QFE phases can vary from 2 to ~3/2 and 5/2 as the two QCPs approach each other with biquadratic and additional lower order couplings, respectively, between the magnetic and FE order parameter fields.

Here, we present results of a comprehensive study of temperature dependence of the dielectric constant (which is essentially the dielectric susceptibility $\chi_e(T)$ in displacive FEs), under zero and non-zero magnetic fields (H), and ac magnetic susceptibility $\chi_m(T)$ in the temperature range 4 to 80K on an M-type hexaferrite SrFe$_{12}$O$_{19}$ (SFO). We show here that the upturn in $\chi_e(T)^{-1}$ reported earlier [49] is due to dipole glass (DG) transition, which is linked to a spin glass (SG) transition via quadratic coupling of the two diverse order parameter fields. The ergodic symmetry breaking temperatures for the dipole glass ($T_{DG}$) and spin glass ($T_{SG}$) transitions nearly coincide within $\pm$ 1K suggesting SFO to exhibit a canonical multiglass state in contrast to previously reported multiglass systems where $T_{DG}$ and $T_{SG}$ differ much more [52–55]. We also show that the dipole glass freezing temperature ($T_{f1}$) increases with magnetic field (H) at a constant frequency suggesting that it is a magnetically stabilised phase [50]. On lowering the temperature further below the DG transition, a re-entrant QPE state appears [50], which can be transformed to another phase, most likely a second dipole glass phase as its freezing temperature ($T_{f2}$) also shifts to higher temperature side with increasing frequency. Unlike the multiglass phase, this transition is not accompanied with any magnetic transition and its freezing temperature $T_{f2}$ decreases with increasing H suggesting that it is destabilised by the field. Finally, we show that near the dipole glass transition temperature for zero magnetic field, $\chi_e(T)^{-1}$ shows a ~$T^{5/2}$ dependence. Such a non-classical exponent is indicative of multiferroic quantum criticality due to the usual biquadratic as well as lower order magnetoelectric couplings of order parameter fields [50]. Evidence for such a coupling is also presented from a study of the variation of dielectric constant with magnetic field at 3 and 20K. These results are discussed in the light of the recent theoretical work on multiferroic quantum criticality [50] and the future directions in which more theoretical efforts are required.

Recent experimental studies suggest that the M-type hexaferrites are uniaxial displacive incipient FEs/AFEs due to the off-centre displacement of $Fe^{3+}$ ions in the trigonal bipyramid (TBP) coordination along the c-axis [32] (see Fig.1(a) and SI for more structural details) as a result of the softening of a zone centre transverse optical (TO) mode whose frequency remains non-zero down to 0K [56] and the lower potential energy [32] at the two symmetrically located off-centre positions schematically depicted in Fig.1(a). Since the $3d^5 Fe^{3+}$ ion in the TBP contributes to both the magnetic moment and the local electric dipole moment, the M-type hexaferrites defy the usual $3d^0$ norm for ferroelectric off-centre displacements in $ABO_3$ perovskites [57]. Fig.1(a) depicts the typical temperature dependence of the real ($\varepsilon'(T)$) and imaginary ($\varepsilon''(T)$) parts of the dielectric constant of $SrFe_{12}O_{19}$ (SFO) in the 4 to 80K range, measured at 150 kHz on a silver-electroded high density and well characterised sintered pellet-piece. The details of sample preparation, characterisations and crystal structure are given in the SI.

The nature of the $\varepsilon_r'(T)$ plot in Fig. 1(a) is similar to $\varepsilon_c'(T)$ plot for single crystals where Barrett's like behaviour along the c-axis and nearly temperature independent behaviour of $\varepsilon_{ab}'(T)$ ($<\varepsilon_c'(T)$) in the *ab*-plane have been reported below 50K [32]. The absolute value of $\varepsilon_r'(T)$ is somewhat reduced in the polycrystalline sample due to the averaging over the $\varepsilon_c'$ and $\varepsilon_{ab}'$ values in randomly oriented grains and slight uncertainty in the geometrical dimensions also. The increasing trend in $\varepsilon_r'(T)$ and $\varepsilon_r''(T)$ above $\sim$60K in Fig. 1(a) is due to conductivity losses, reported in single crystals also, as SFO is an n-type semiconductor with a rather small bandgap [58]. The Barrett's law [29] gives fairly satisfactory fit to the observed $\varepsilon_r'(T)$ below 50K with M =(1.59±0.01), $T_0$ =(44.89±0.3)K, and $T_1$ =(108.01±0.1)K, as can be seen from the continuous line in Fig.1(a). The positive value of $T_0$ indicates the presence of ferroelectric like correlations unlike the AFE correlations with negative $T_0$ reported for the isostructural BFO [22,32]. Although the value of these parameters naturally differ somewhat from those corresponding to the single crystal values [32], they do indicate the QPE behaviour at lower temperatures.

The first indication of something is amiss in the Barrett's fit comes from the temperature dependence of the imaginary part of the dielectric constant $\varepsilon_r''(T)$, which surprisingly has not been given in the earlier reports [32,49]. The $\varepsilon_r''(T)$ plot in Fig.1(a) shows a peak around $T_{f1}'' \sim 12.8K$ related to a tiny peak in the $\varepsilon_r'(T)$ plot at $T_{f1}' \sim 13.4K$. Our $T_{f1}'$ is close to the upturn temperature of $\sim 14$ K in $\chi_e(T)^{-1}$ reported earlier [49]. The second peak in the $\varepsilon_r''(T)$ plot at $T_{f1}'' \sim 24K$ does not lead to any obvious anomaly in the $\varepsilon_r'(T)$ plot. The tiny peak at $\sim 13.4K$ for 150 kHz is not accounted for by Barrett's law. It precedes the onset of a nearly temperature independent plateau region of the $\varepsilon'(T)$ plot at lower temperatures indicating the emergence of a reentrant QPE state of SFO predicted theoretically for multiferroic quantum criticality [50]. It is worth to note that the peak in the $\varepsilon_r''(T)$ plot occurs at a slightly lower temperature ($T_{f1}''$=12.8K) than the peak in $\varepsilon_r'(T)$ ($T_{f1}'$=13.4K) in Fig.1(a). This is unlike that for a regular polar transition, for which $T_{f1}'$ must be equal to $T_{f1}''$ as per the Kramer-Kronig relationship [26]. For DG (and relaxor ferroelectric (RFE)) transitions, on the other hand, $T_{f1}''$ is known to be less than $T_{f1}'$ [22,59]. The frequency dependent $\varepsilon_r'(T)$ response, measured over 4 to 40K temperature range, reveals that $T_{f1}'$ shifts to higher temperatures with increasing value of the measuring frequency ($\omega$) (see Fig.2) similar to DG [60,61] and RFE [59,62,63] transitions. The temperature dependence of the dipolar relaxation time ($\tau$), obtained from $\omega\tau=1$ relationship for each temperature $T_f'(\omega)$ [62,63], shows non-Arrhenius behaviour in the $\ln(\tau)$ versus $1/T_f'$ plot (see inset "a" of Fig.2). This is a typical characteristic of DGs [60,61] and RFEs [59,62,63], where the temperature dependence of the relaxation time $\tau$ has been analysed in terms of Vogel-Fulcher (VF) [59,62] $\tau$ =$\tau_0 \exp(E_a/k_B(T-T_{VF}))$ and power-law $\tau = \tau_0((T_f -T_{DG})/T_{DG})^{z\upsilon}$ [52] behaviours, in analogy with the SG systems [64]. Here $\tau_0$ is the inverse of the attempt frequency, $E_a$ is activation energy for the dipole relaxation process, $k_B$ is the Boltzmann constant, $T_{VF}$ and $T_{DG}$ are characteristics temperatures at which the slowest polar dynamics diverges signalling ergodic symmetry breaking and $z\nu$ is the dynamical critical exponent related to the correlation length [64]. The fit between the observed and calculated temperature dependence of $\tau$ for both VF and power laws are quite satisfactory, as can be seen from the insets (a) and (b) of Fig.2, respectively, with the fitting parameters $z\nu$ =(2.04±0.01), $\tau_0$ =(4.7±0.2) x $10^{-7}$ s and $T_{DG}$ =(12.05±0.01)K for the power law and $E_a$=0.46eV, $\tau_0$ =(2.5±0.1)x$10^{-6}$ s and $T_{VF}$ =(11.48±0.01)K for the VF law. The relatively large value of $\tau_0$ indicates the presence of clusters of correlated dipoles with frustrated inter-cluster interactions similar to cluster spin glasses [64]. Further, the ergodic symmetry breaking temperatures for the DG state for the two laws are very close to each other with $T_{DG} \sim$12K $T_{VF} \sim$11.5K.

The peak temperatures $T'_{f1}$ (150kHz) =13.4K and $T''_{f2}$ (150kHz) =24K in the $\varepsilon_r$'(T) and $\varepsilon_r$''(T) plots correlate well with the onset of two relaxation steps at $\sim$13.4K and $\sim$24K in the ac magnetic susceptibility $\chi_m$'(T) plot shown in Fig.1(b), another key measurement missing in the previous studies [32,49]. Interestingly, even the hump in the $\varepsilon$''(T) plot around 55K correlates well with a weak magnetic anomaly shown more clearly in the inset of Fig1(b). We find that similar to $BaFe_{12}O_{19}$ [65–67], SFO undergoes a succession of SG transitions below room temperature but the details of all these transitions is outside the scope of the present report. We shall focus only on the lowest temperature relaxation step in $\chi_m$'(T) plot shown in Fig 1(b) which correlates with the DG transition in the $\varepsilon_r$'($\omega$,T) plot of Fig.2. Since it was not possible to extract the information about SG freezing temperatures $T'_f(\omega)$ from the $\chi_m$'(T) plot (see Figs.1(b)) reliably, we analysed the frequency dependence of the upturn temperature in $\chi_m$''(T), which is known to coincide with $T'_f(\omega)$ for SG systems [64,68], as can be seen in Fig.1(b) also. This is because the onset of increase in dissipation revealed by $\chi_m$''(T) in glassy systems coincides with the onset of glassy freezing at $T_f'$ revealed by the peak in $\chi_m'(T)$ plot. The frequency dependence of the upturn temperature in the $\chi''(T)$ plot is shown in Fig.3(a) and the results of its analysis in terms of power law and VF relationships are shown in Fig.3(b) and its inset, respectively. Both the analysis confirm the existence of a critical temperature, $T_{SG}$ =(11.05±0.01)K and $T_{VF}$ =(10.66±0.02)K, at which the ergodic symmetry is broken and τ diverges for the fitting parameters: zυ =(0.84±0.02), $\tau_0$ =(4.7±0.3)x$10^{-3}$ s and $E_a$ =0.11meV, $\tau_0$ =(1.25±0.05)x$10^{-2}$ s, respectively. Interestingly, the ergodicity breaking temperatures for the DG and SG transitions nearly coincide ($T_{DG}$∼$T_{SG}$) within ±1K. All these results indicate for the first time that $SrFe_{12}O_{19}$ exhibits a multiglass state [52–55], preceding the onset of the QPE saturation of ε'(T).

Both the multiglass and the re-entrant QPE states of $SrFe_{12}O_{19}$ are sensitive to perturbations caused by dc magnetic field (H) as can be seen from Fig.4(a) which depicts $\varepsilon_r'(T)$ plots measured at 150 kHz under various H in the range 100 to 30000 Oe. As is evident from this figure, even a small magnetic field of 100 Oe is sufficient to suppress the quantum fluctuations in the re-entrant QPE state and stabilise a lower temperature phase giving rise to a second peak in $\varepsilon_r'(T)$ at ∼7.9K. The transition temperature corresponding to this phase is found to decrease non-monotonically with increasing magnetic field while the multiglass freezing temperature around 13.4K increases non-monotonically with increasing field, as can be seen from Fig.4(a) and (b) and the inset of (b). It is interesting to note that the DG character of the higher temperature transition around 13.4K is retained under dc magnetic field bias as it continues to show the characteristic shift in $T_f'(\omega)$ towards higher temperature side on increasing the frequency (see Fig.4(c)) with $T_{DG}$ ∼(12.88±0.01)K and $T_{VF}$∼(12.42±0.02)K (see Fig.4(d) and its inset) for the power-law and VF fits, respectively. The frequency dispersion in the $T_f'(\omega)$ of the lower temperature transition points towards another DG transition but it could not be confirmed due to uncertainty in location of the peak temperatures.

As said earlier, the tiny peak in the $\varepsilon'(T)$ plot in Fig. 1(a) and Fig. 2 has been previously analysed [49] in terms of quantum critical model for uniaxial ferroelectrics [44] in the 20 to 35K range [49]. Since there is no estimate of the non-zero gap at the zone centre for the soft TO mode of SFO [56], the lower bound of 20K is somewhat tentative in ref. [49] . More significantly, since we are dealing with a  multiferroic multiglass transition, which can be further stabilised by magnetic field as evidenced by the increase in $T_f'$ shown in Fig.4(b), the multiferroic quantum critical models are required [50,51]. The presence of re-entrant QPE state in SFO below a magnetically enhanced DG state is consistent with the predictions of the multiferroic quantum criticality [50]. The temperature dependence of the inverse of the dielectric constant ($\varepsilon_r'(T)^{-1}$), which is essentially the dielectric susceptibility($\chi_e(T)^{-1}$), was analysed in terms of $T^n$ type dependence in the QCR. The exponent n can be 1, 2, 3, 3/2 and ~5/2 for the classical FE C-W law, quantum critical multiaxial FE, quantum critical uniaxial FE, multiferroic quantum critical models with FE and magnetic QCPs coinciding with each other having biquadratic coupling and multiferroic with additional lower order couplings, respectively. The best fit is obtained for n ~5/2 which covers the entire curvature of $\varepsilon_r'(T)^{-1}$ curve in the temperature range 15 to 36K, as can be seen from Fig.5(a) (compare with Figs.S2(a-e) of SI). This exponent is obviously a non-classical exponent and may involve the effect of biquadratic and lower order magnetoelectric couplings in SFO. To explore the nature of order parameter field couplings in SFO,  the dc magnetisation (M), $\varepsilon'$ and dielectric loss tangent (tan$\delta$)  were measured as a function of field at 20 and 3K following the approach discussed in the literature [69,70]. The percentage change in the dielectric constant $\varepsilon'(H)$ with respect to $\varepsilon'(H=0)$, i.e., ($\%\Delta\varepsilon'/\varepsilon'(H=0) = 100(\varepsilon'(H)-\varepsilon'(H=0))/\varepsilon'(H=0)$, increases with H prior to the saturation of magnetization but begins to decrease with further increase in H in the saturation region showing positive and negative magnetodielectric couplings, respectively (see Fig.S3 and section S4 of SI). The magnetic field independence of the $\tan(\delta)$ in this figure shows that the magnetodielectric coupling is not due to magnetoresistance effects, as per the criterion outlined in ref. [71]. The maximum change in the dielectric constant $\varepsilon_r'(H)$ due to the applied magnetic field is ~0.06% which is nearly half of that in multiferroic QPEs like $EuTiO_3$ [72] but is comparable to that in $NiCr_2O_4$ and $Mn_3O_4$ [70,73]. While the variation of $\%\Delta\varepsilon'/\varepsilon'(H=0)$ for SFO shows $M^2$ dependence at higher fields, consistent with the predictions for biquadratic order parameter couplings (ref. [69] and see Fig.5(b)), it also shows departure from $M^2$ dependence at lower fields possibly due to the presence of lower order couplings (see Fig.S4 and section S4 of SI). We believe that the magnetodielectric coupling is responsible for the appearance of the multiglass state of SFO.

To summarise, we have presented evidence for a re-entrant QPE state preceded by a dipole glass transition with a non-classical exponent (~5/2) in the quantum critical regime of $SrFe_{12}O_{19}$. The dipole glass (DG) transition is accompanied with a spin glass transition characteristic of a canonical multiglass state in which the two diverse ergodic symmetry breaking temperatures nearly coincide ($T_{DG} \sim T_{SG}$) within ±1K due to magnetodielectric coupling. Further, we have shown that application of external magnetic field can induce another transition, most likely a second DG transition, in the re-entrant QPE state. This lower temperature transition is different from the higher temperature multiglass transition as it is not accompanied with any magnetic transition. Since its transition temperature decreases with increasing magnetic field, it shows the possibility of locating a quantum critical point (QCP) in $SrFe_{12}O_{19}$ at much higher magnetic fields than that used in the present work. The freezing temperature of the higher temperature dipole glass state, on the other hand, increases with the magnetic field, i.e., it is magnetically enhanced. While the observation of re-entrant QPE state and a magnetically enhanced DG state bears broad phenomenological similarities with the theoretical predictions for the multiferroic quantum critical model [50], there is obviously a need for developing a quantum critical model for a multiglass state with nearly coincident critical ergodic symmetry breaking temperatures ($T_{DG} \sim T_{SG}$). We believe our findings will stimulate further experimental and theoretical studies in multiferroic glassy systems.

**Acknowledgements:** Financial support by the DST provided within the framework of the India@DESY collaboration is gratefully acknowledged. The authors acknowledge DESY (Hamburg, Germany), a member of the Helmholtz Association HGF, for the provision of experimental facilities. KK is grateful to Prof Premkumar Senguttuvan from JNCASR, Bengaluru for his support and encouragement.

**Figure caption**

**Figure 01: (a)** Plots of $\varepsilon'(T)$ and $\varepsilon''(T)$ at 150kHz. **(b)** Plots of $\chi_m'(T)$ and $\chi_m''(T)$ at 5kHz. Solid black line through the $\varepsilon'(T)$ data points in **(a)** corresponds to the Barret's fit. Inset of **(a)** shows TBP polyhedra with the two off-centre positions (4e) of $Fe^{3+}$.

**Figure 02:** Plot of $\varepsilon'(T)$ at various frequencies: ■:1kHz, ●:30kHz, ▲:70kHz, ▼:100kHz, ◆:150kHz & ◀:200kHz). Insets show **(a)** V-F fit and **(b)** power law fit for the relaxation time $\tau$.

**Figure 03: (a)** Plot of $\chi_m''(T)$ at various frequencies: ■:50Hz, ▲:75Hz, ◆:100Hz, ▶:150Hz, ★:200Hz. For clarity each curve is shifted by a constant value 8 x $10^{-5}$. Black solid line through the data points is guide to the eyes. **(b)** The power law and VF law fits (inset) for the relaxation time $\tau$ at each $T_f$ obtained from the upturn point marked with an arrow in **(a)**.

**Figure 04: (a)** Variation of capacitance $C'_P$ with T at 150kHz measured under dc magnetic fields, where the arrows passing through the freezing temperatures $T_{f1}$ and $T_{f2}$ are guides to the eyes: H=▲:100Oe, ▼:300Oe, ◆:500Oe, ▶:1000Oe, ◀:5000Oe, ●:10000Oe, ★:30000Oe. **(b)** Variation of $T_{f1}$ and $T_{f2}$ with field (H). **(c)** $C'_P(\omega, T)$ measured at various frequencies at H=100 Oe. **(d)** Power law fit with $T_{DG} \sim (12.88\pm0.01)$K, $z\upsilon =(0.32\pm0.01)$, and $\tau_0 =(1.07\pm0.05)$ x $10^{-5}$ s. Inset of **(d)** shows V-F law fit with $T_{VF} \sim (12.42\pm0.02)$K, $\tau_0 =(2.41\pm0.04)$ x $10^{-5}$ s and $E_a =0.061$meV

**Figure 05: (a)** Inverse dielectric susceptibility $(1/\chi_e)$ vs T with fits for various critical exponents n = 1,2,3, 3/2 and 5/2. **(b)** The percentage change in $\varepsilon'$ under DC magnetic field bias versus $M^2$ plots at 3K and 20K.

**Figures**

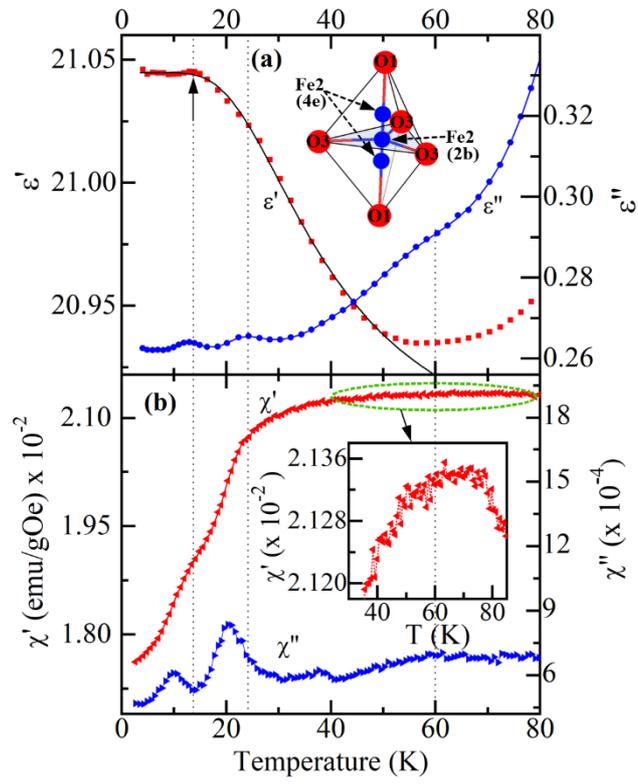

**Figure 01**

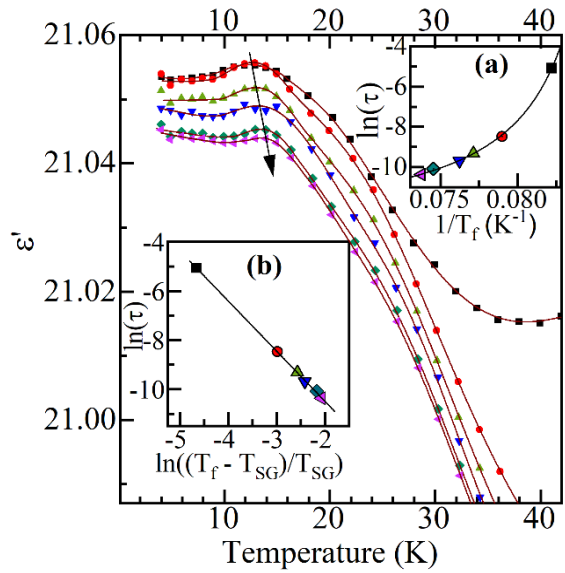

**Figure 02**

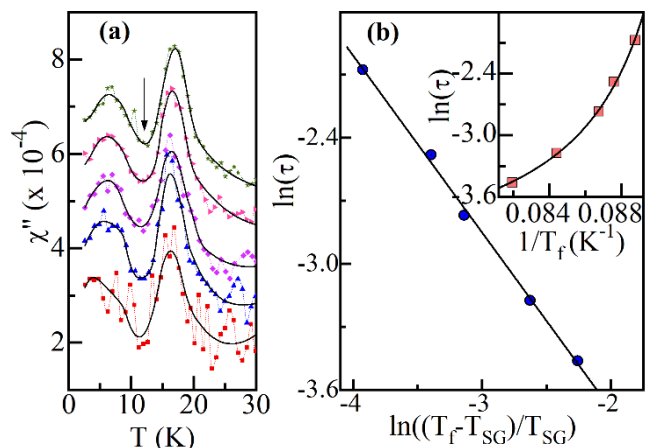

**Figure 03**

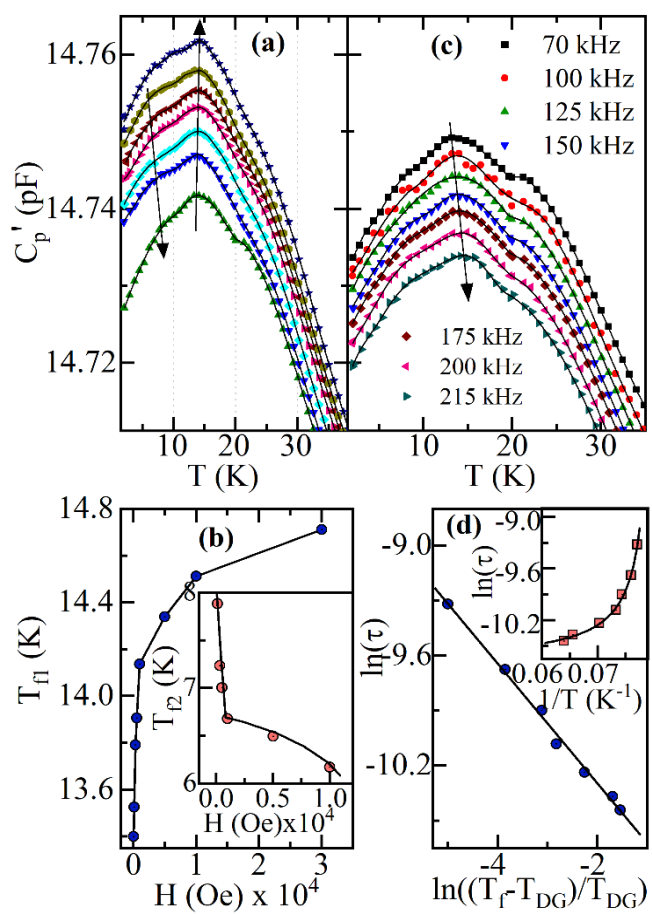

**Figure 04**

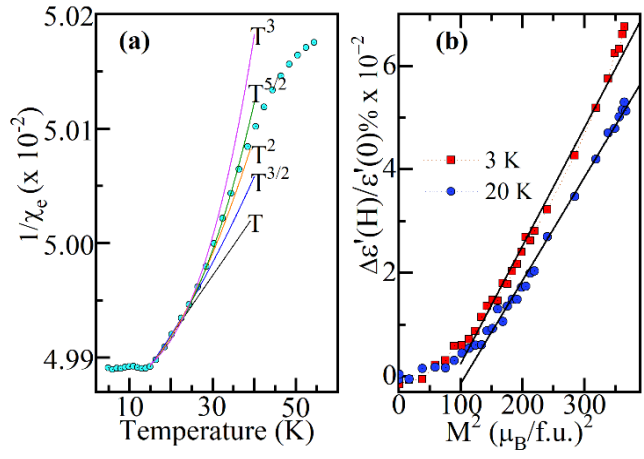

**Figure 05**